\documentclass{aastex63}
\usepackage{graphicx}
\usepackage{hyperref}
\usepackage{amsmath}

\accepted{February 9, 2023}
\submitjournal{ApJ}

\shorttitle{HD 53143's peculiar disk}
\shortauthors{Stark et al.}

\begin{document}

\title{The apparent absence of forward scattering in the HD 53143 debris disk}

\correspondingauthor{Christopher Stark}
\email{christopher.c.stark@nasa.gov}

\correspondingauthor{Bin Ren}
\email{bin.ren@oca.eu}

\author{Christopher C. Stark}
\affiliation{NASA Goddard Space Flight Center, Exoplanets and Stellar Astrophysics Laboratory, Code 667, Greenbelt, MD 20771, USA}

\author{Bin Ren}
\affiliation{Universit\'{e} C\^{o}te d'Azur, Observatoire de la C\^{o}te d'Azur, CNRS, Laboratoire Lagrange, F-06304 Nice, France}
\affiliation{Universit\'{e} Grenoble Alpes, CNRS, Institut de Plan\'{e}tologie et d'Astrophysique (IPAG), F-38000 Grenoble, France}
\affiliation{Department of Astronomy, California Institute of Technology, MC 249-17, 1200 East California Boulevard, Pasadena, CA 91125, USA}

\author{Meredith A. MacGregor}
\author{Ward S. Howard}
\affiliation{Department of Astrophysical and Planetary Sciences, University of Colorado, 2000 Colorado Avenue, Boulder, CO 80309, USA}

\author{Spencer A. Hurt}
\affiliation{Department of Earth Sciences, University of Oregon, Eugene, OR 97403, USA}

\author{Alycia J. Weinberger}
\affiliation{Earth \& Planets Laboratory, Carnegie Institution for Science, 5241 Broad Branch Road NW, Washington, DC 20015, USA}

\author{Glenn Schneider}
\affiliation{Steward Observatory, The University of Arizona, 933 North Cherry Avenue, Tucson, AZ 85721, USA}

\author{Elodie Choquet}
\affiliation{Aix Marseille Univ, CNRS, CNES, LAM, Marseille, France}

\begin{abstract}	

HD 53143 is a mature Sun-like star and host to a broad disk of dusty debris, including a cold outer ring of planetesimals near 90 AU. Unlike most other inclined debris disks imaged at visible wavelengths, the cold disk around HD 53143 appears as disconnected ``arcs" of material, with no forward scattering side detected to date. We present new, deeper \emph{Hubble Space Telescope} (HST) Space Telescope Imaging Spectrograph (STIS) coronagraphic observations of the HD 53143 debris disk and show that the forward scattering side of the disk remains undetected. By fitting our KLIP-reduced observations via forward modeling with an optically thin disk model, we show that fitting the visible wavelength images with an azimuthally symmetric disk with unconstrained orientation results in an unphysical edge-on orientation that is at odds with recent ALMA observations, while constraining the orientation to that observed by ALMA results in nearly isotropically scattering dust. We show that the HD53143 host star exhibits significant stellar variations due to spot rotation and revisit age estimates for this system.

\end{abstract}


\section{Introduction} \label{sec:intro}

The outer regions of extrasolar planetary systems are commonly thought to harbor planetesimals, analogous to the Kuiper Belt in the Solar System. The most massive of these systems, which are currently accessible with imaging techniques from visible to millimeter wavelengths, exhibit disks of dusty debris thought to be continually generated by these planetesimals. At millimeter wavelengths, observations are sensitive to larger dust grains that are relatively unaffected by radiative forces and thought to track the orbits of their larger parent bodies \citep[e.g.,][]{hughes2018}. Debris disks often appear as well-defined circumstellar belts/rings at millimeter wavelengths, suggesting parent bodies are relatively confined in semi-major axis \citep[e.g.,][]{macgregor2019}. At visible wavelengths, observations are sensitive to sub-micron-sized dust grains, which for high luminosity stars are likely expelled from the system due to radiation pressure, but may remain bound for stars with similar luminosity to HD 53143 \citep[e.g.,][]{arnold2019}. At near-infrared wavelengths, observations probe micron-sized dust grains, which remain bound but are sent onto eccentric orbits with large semi-major axes by radiation pressure \citep[e.g.,][]{wyatt1999}. The majority, but not all, of these small marginally bound grains are collisionally destroyed prior to being dragged into the inner system by Poynting-Robertson and corpuscular drag \citep{wyatt2005,kennedy2015,rigley2020}. The result is a ring with a well-defined inner edge and extended outer halo at visible wavelengths \citep[e.g.,][]{thebault2014}. Additionally, inclined disks often exhibit a brightness asymmetry along the minor axis at visible wavelengths, with the near side of the disk appearing brighter due to the forward scattering of starlight by small dust grains \citep{augereau1999,weinberger1999}.

HD 53143 hosts a peculiar disk that appears to depart from the typical appearance of debris disks. HD 53143 is a Sun-like G9V star at 18.3 pc with an estimated age of 1 Gyr \citep{kalas2006}. This mature star hosts an optically thin debris disk with $L_{\rm disk}/L_{\star} = 0.025$\% \citep[]{zuckerman2004}, fairly bright at infrared wavelengths for a Gyr-old system. The debris disk around HD 53143 extends to the warm inner regions, with SED model fits suggesting dust in the 0.6 -- 16 AU region \citep{chen2006,chen2014}. The outer cold disk was imaged coronagraphically for the first time by \cite{kalas2006} using the Advanced Camera for Surveys (ACS) High Resolution Camera (HRC) on HST. With a 1.8\arcsec occulting spot, the HST ACS images revealed faint arc-like features along the NW and SE quadrants, extending from 55 to 110 AU. The region interior to 3\arcsec was dominated by PSF residuals, and the arcs were interpreted as the ansae of a disk inclined by $\sim45^{\circ}$ with a PA$=142\pm2^{\circ}$. 

The HD 53143 debris disk was observed again by \cite{schneider2014} using the HST STIS broadband coronagraphs (note it is incorrectly referred to as HD 53154 sporadically throughout the manuscript). \cite{schneider2014} used the WedgeA-0.6 and WedgeA-1.0 masks to image the region of the disk exterior to its inner working angle (IWA) of 0.3\arcsec. The STIS observations revealed azimuthally symmetric flux interior to 50 AU that appeared more like a face-on inner disk than a disk aligned with the outer arcs. 

Notably, in both observations, there was no observed flux from the disk connecting the arcs of material. The outer arcs at 90 AU appear detached from the rest of the disk (yet still comoving with the star). If these arcs are the ansae of a ring-like disk, we would expect to see even brighter disk flux along the minor axis due to the forward scattering that is typical of other debris disks.

\cite{macgregor2022} observed the disk with ALMA Band 6 at 1.36 mm. These observations revealed the ring-like structure of the outer disk for the first time, showing that the larger grains, and presumably the parent planetesimals, are confined to a $\sim20$ AU wide belt with semi-major axis $90.1\pm0.5$ AU, inclined by $56.2\pm0.4^{\circ}$ from face-on, and with a PA$=157.3\pm0.3^{\circ}$. With a measured eccentricity of $0.21\pm0.02$, the outer belt of material is the most eccentric debris belt observed to date. Model fits to the belt suggest apocenter roughly aligns with the NW ansa, such that an observed enhancement in mm flux at this location can be explained by an apocenter glow effect previously observed in the eccentric Fomalhaut belt \citep{macgregor2017}. Like the STIS observations, ALMA also revealed excess flux near the star, though the bi-lobed structured was more suggestive of an edge-on inner disk than the face-on disk inferred from STIS observations. Notably, \cite{macgregor2022} found the stellar flux to vary significantly during the multi-epoch ALMA observations that spanned $\sim12$ days, which they attribute to stellar flares from this mature Sun-like star.

With these peculiarities in mind, we observed HD 53143 with the HST STIS WedgeA-1.8 coronagraph to peer deeper into the outer regions of this system in an effort to constrain the geometry of the arcs to the NW and SE, as well as search for signs of the forward scattering side of the ring structure at visible wavelengths that should connect the ansae. In Sections \ref{sec:observations} and \ref{sec:datareduction} we describe our observations and data reduction methods. In Sections \ref{sec:analysis} and \ref{sec:discussion} we present model fits to the disk and discuss our interpretation of this system.

\section{Observations} \label{sec:observations}

We observed the HD 53143 debris disk via STIS coronagraphy using the WedgeA-1.8 occulting mask to focus on the outer arcs at $\sim$5\arcsec. We chose WedgeA-1.8 to reduce stellar residuals and improve $S/N$ in the region of the arcs, while maintaining adequate roll coverage and access to a large library of previous PSFs in the event that principle component analysis (PCA) was needed for data reduction. Table \ref{table:observations} summarizes our observations. We observed HD 53143 for a total of 15 orbits, split into three groups of five orbits to improve schedulability. Each group consisted of a non-interruptible sequence interleaving the PSF reference star in the middle to minimize time-dependent artifacts due to changes in the telescope's optical assembly temperature.  We varied the orientation of the observations over $\sim18^{\circ}$ to observe regions occulted by the wedge at some angles and obtain broader coverage of the outer arcs. We avoided full roll coverage in favor of deeper exposure times on the arcs, leaving unobserved areas where the occulting mask aligned with the disk's minor axis. We designed individual exposures such that the stellar PSF residuals reached 90\% of the full-well depth of the STIS CCD, based on by prior observations by \cite{schneider2014}.

\begin{deluxetable}{cccc}
\tablewidth{0pt}
\footnotesize
\tablecaption{Observation Log\label{table:observations}}
\tablehead{
\colhead{Target} & \colhead{UT Obs. Date} & \colhead{Orient} & \colhead{Exposure Time}\\
\colhead{} & \colhead{} & \colhead{($^{\circ}$)} & \colhead{(s)}\\
}
\startdata
HD 53143 & 2021 Oct 11 & $-128.9^{\circ}$ & 2400\\
HD 53143 & 2021 Oct 11 & $-130.4^{\circ}$ & 2400\\
HD 58895 & 2021 Oct 11 & $-135.3^{\circ}$ & 2340\\
HD 53143 & 2021 Oct 11 & $-131.9^{\circ}$ & 2400\\
HD 53143 & 2021 Oct 11 & $-133.4^{\circ}$ & 2400\\
\hline
HD 53143 & 2021 Oct 14 & $-122.9^{\circ}$ & 2400\\
HD 53143 & 2021 Oct 14 & $-124.4^{\circ}$ & 2400\\
HD 58895 & 2021 Oct 14 & $-127.9^{\circ}$ & 2340\\
HD 53143 & 2021 Oct 14 & $-125.9^{\circ}$ & 2400\\
HD 53143 & 2021 Oct 14 & $-127.4^{\circ}$ & 2400\\
\hline
HD 53143 & 2021 Oct 22 & $-116.9^{\circ}$ & 2300\\
HD 53143 & 2021 Oct 22 & $-118.4^{\circ}$ & 2300\\
HD 58895 & 2021 Oct 22 & $-123.6^{\circ}$ & 2240\\
HD 53143 & 2021 Oct 23 & $-119.9^{\circ}$ & 2300\\
HD 53143 & 2021 Oct 23 & $-121.4^{\circ}$ & 2300\\
\enddata
\vspace{-0.1in}
\end{deluxetable}

Because the STIS coronagraph has a broad bandpass (0.2–1.2 $\mu$m) and diffracted residual starlight varies with wavelength, care must be taken when choosing a reference PSF star. The reference star used by \cite{schneider2014} was HD 59780, which has $(B-V)=0.95$. This compares relatively well with HD 53143's $(B-V)=0.81$ for an expected $\Delta(B-V)=-0.14$. At the time of our observations, the circularly symmetric flux observed by \cite{schneider2014} was suspected to be chromatic residual starlight due to a color mismatch between target and reference star. Thus, we chose a PSF reference star with similar $\Delta(B-V)$ magnitude, but opposite sign. We adopted HD 58895 as our PSF reference star, which has $(B-V)=-0.70$ for an expected color difference of $\Delta(B-V)=0.11$. In spite of this, as discussed below, the observed mismatch in PSFs was significantly larger than expected, requiring us to form a reference PSF from the existing library of STIS WedgeA-1.8 observations via PCA techniques. We discuss possible causes for this mismatch in Section \ref{sec:discussion}.

\section{Data Reduction} \label{sec:datareduction}

Intra-visit pointing stability was better than 0.2 pixels end-to-end RMS. A small ($<10$ mas) systematic drift in the observations was noticed in the $(+x,-y)$ direction in the Science Instrument Aperture File (SIAF) frame, most likely due to the guidance system. Neither the jitter nor drift is expected to significantly impact our results. The astrophysical background varied significantly between visits, likely due to changes in stray light from Earth, and had to be subtracted on a visit-by-visit basis. 

\begin{figure}[h]
	\centering	
	\includegraphics[width=4in]{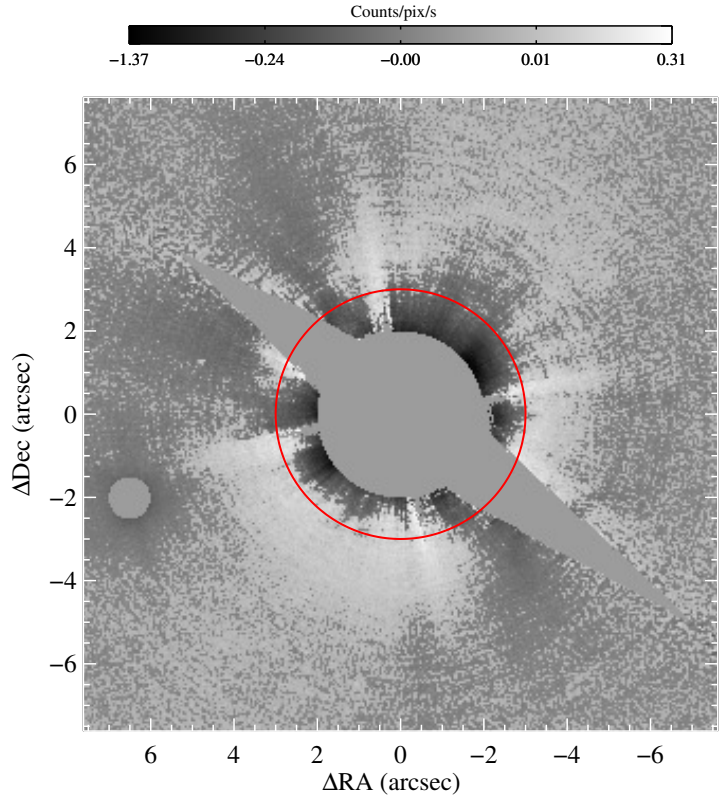}
	\caption{Reduction of our STIS observations using classical PSF subtraction, shown with a circle of radius 3\arcsec for reference. The inner 2\arcsec (as well as a background object to the SE) has been masked off due to residuals from the poor match of our PSF reference star. The NW and SE arcs at 5\arcsec are detected, but their extent and geometry is difficult to discern amongst PSF subtraction artifacts.\label{fig:classicalpsfsubtraction}}
\end{figure}

Figure \ref{fig:classicalpsfsubtraction} shows the result of classical PSF subtraction using our observed PSF reference star HD 58895. The red circle has radius 3\arcsec~for reference. The stellar residuals are much larger than anticipated and overwhelm the arcs we seek to observe at 5\arcsec, in spite of a smaller absolute $\Delta(B-V)$ than prior observations.

\begin{figure}[h]
	\centering	
	\includegraphics[width=4in]{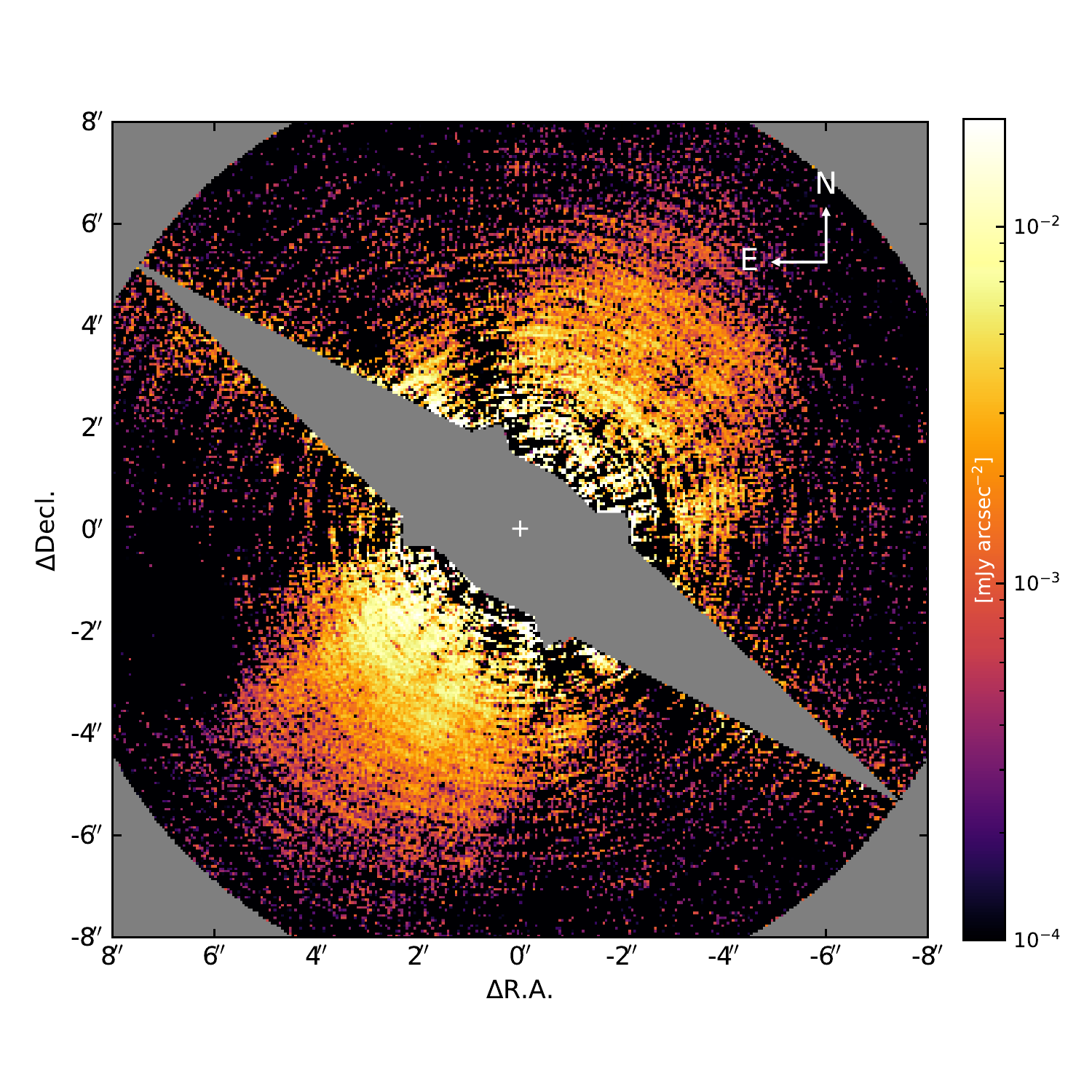}
	\caption{Reduction of our STIS observations using the KLIP method, clearly revealing the NW and SE arcs. Image units are mJy arcsec$^{-2}$.\label{fig:klipreduction}}
\end{figure}

Given the poor match of our PSF reference star, we opted to reduce the data using a PCA analysis via the KLIP algorithm \citep{soummer2012}. We assembled the reference archive from \citet{ren2017} to capture the variation of the point spread function (PSF) for the STIS broadband filter. There are 274 individual readout images in the STIS WedgeA1.8 location that are thought to host no circumstellar structure. For each HD 53143 readout, we selected 30 of the highest-correlated references and performed KLIP subtraction of the PSF using all of the KL components. We rotated all of the 96 exposures to north-up and east-left, then obtained their median as our reduction result for the HD 53143 system. Figure~\ref{fig:klipreduction} shows the final reduced image. The NW and SE arcs of the outer disk are clearly visible.

We note that we attempted to use the non-negative matrix factorization (NMF) method of \citet{ren2018}. The spatial extent of the HD 53143 disk limits the applicability of this method and makes it computationally challenging. Further, the smaller availability of WedgeA-1.8 PSFs, combined with apparent color variability (see Section \ref{psfvariability}) degraded the NMF results.

\section{Analysis} \label{sec:analysis}

Despite significantly increased exposure time, there remains no sign of a ring-like structure connecting the NW and SE arcs at visible wavelengths. This is notable given that one side of this ring should be forward-scattering, and thus significantly brighter than the arcs.

In an attempt to better understand the observed scattered light distribution, we fit models of an optically thin disk to the observed data set, the first time this has been attempted for the HD 53143 system. We adopted a radial profile for the surface brightness based on the \citet{augereau1999} double power-law profile. In the direction perpendicular to the midplane, we again used the distribution of \citet{augereau1999} and set the parameters $\beta=1$ and $\gamma=2$, such that our distribution was equivalent to a Gaussian dispersion of scale height $h=H/r$, where $r$ is circumstellar distance measured in the midplane and $H$ is the height above the mid-plane. We modeled the scattering phase function of the disk using a linear combination of two Henyey-Greenstein (HG) scattering phase functions (SPF) with scattering asymmetry parameters $g_1$ and $g_2$, and weighting factors $f_{g_1}$ and $f_{g_2}=1-f_{g_1}$ \citep{henyey1941}.  

The KLIP PSF subtraction algorithm is known to overfit data, which could cause a reduction in surface brightness and a morphology change \citep[e.g.,][]{soummer2012}. To recover the brightness and morphology of the HD 53143 disk, we adopted a forward modeling approach.  We subtracted each modeled disk from the original readouts, performed KLIP reduction following an identical approach for the reduction of the original readouts, and inspected the KLIP reduction residuals. To find the best-fit model, we maximized the likelihood function assuming pixels are independent and follow Gaussian statistics, using the same procedure as in \citep{ren2021}. We performed the forward modeling procedure by exploring a broad range of parameters using the emcee package \citep{foremanmackey2013}. We implemented the computation on a computer cluster using the DebrisDiskFM package \citep{ren2019} to enable efficient real-time calculation of the model parameters. To make the multi-parameter fitting process numerically tractable, we binned the STIS images on a 3$\times$3 pixel basis.

We adopted uniform priors for all nine parameters of our MCMC analysis. We allowed parameters to vary over the ranges listed in Table \ref{table:mcmc_parameters}. For initial guesses, we sampled each of the parameters uniformly 18 times over a smaller range, also provided in Table \ref{table:mcmc_parameters}. We adopted 18 walkers, 16k MCMC steps, and 8k burn-in steps, after which the likelihood values were not systematically increasing.

\begin{deluxetable}{lccc}
\tablewidth{0pt}
\footnotesize
\tablecaption{Disk Model Parameter Ranges for MCMC Analysis\label{table:mcmc_parameters}}
\tablehead{
\colhead{Parameter} & \colhead{Range Allowed} & \colhead{Initial Guess Range}\\
}
\startdata
Position angle (PA) & $90 < {\rm PA} < 180^{\circ}$ & $130 < {\rm PA} < 170^{\circ}$ \\
Inclination ($i$) & $45 < i < 90^{\circ}$ & $70 < i < 85^{\circ}$ \\
Cross-over radius ($R_c$) & $50 < R_c < 280$ AU & $70 < R_c < 140$ AU \\
Inner power law ($\alpha_{\rm in}$) & $0 < \alpha_{\rm in} < 10$ & $2 < \alpha_{\rm in} < 9$ \\
Outer power law ($\alpha_{\rm out}$) & $-10 < \alpha_{\rm out} < 0$ & $-3 < \alpha_{\rm out} < -1$ \\
Scale height ($h$) & $0 < h < 0.5$ & $0.04 < h < 0.2$ \\
Forward scattering parameter ($g_1$) & $0 < g_1 < 1$ & $0.2 < g_1 < 0.8$ \\
Back-scattering parameter ($g_2$) & $-1 < g_2 < 0$ & $-0.8 < g_2 < 0$ \\
Scattering weight ($f_{g_1}$) & $0 < f_{g_1} < 1$ & $0.7 < f_{g_1} < 1$ \\
\enddata
\vspace{-0.1in}
\end{deluxetable}



\begin{figure}[h]
	\centering	
	\includegraphics[width=7in]{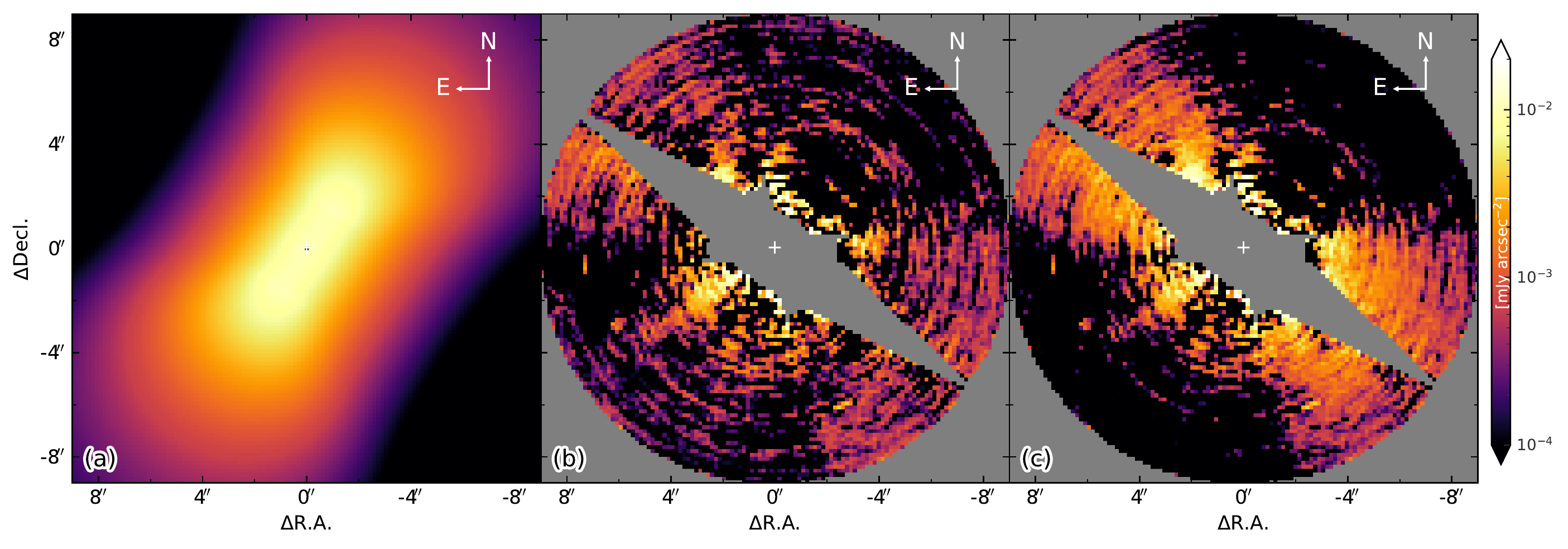}
	\caption{\emph{Left:} Best fit model to the observations. The fitting routine avoids placing flux along the minor axis of the ALMA observed ring by favoring an edge-on orientation. \emph{Middle:} KLIP-reduced data set with the best fit model subtracted. \emph{Right:} The model-subtracted residuals with the best fit model injected at a $\Delta$PA$=90^{\circ}$, showing that flux to the NE and SW due to forward scattering would have been detectable if it were present. \label{fig:bestfitmodel}}
\end{figure}

\begin{figure}[h]
	\centering	
	\includegraphics[width=7in]{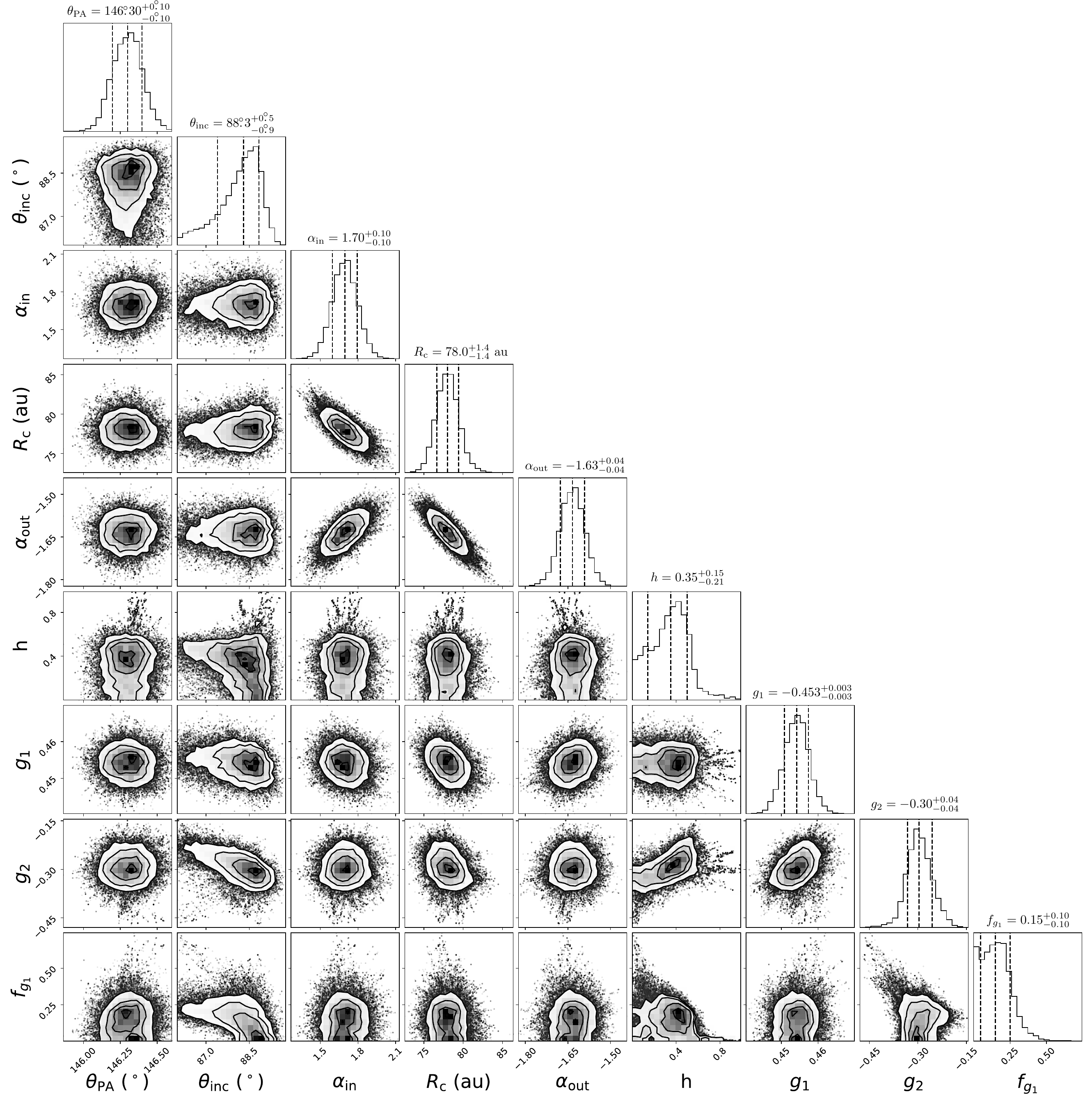}
	\caption{Posterior distribution for our best fit disk model.\label{fig:posteriors}}
\end{figure}

The left panel of Figure \ref{fig:bestfitmodel} shows our best fit model when all parameters are freely explored. The best fit model has a PA$=146.3\pm0.1^{\circ}$, which agrees with prior visible wavelength estimates \citep{kalas2006,schneider2014} but disagrees with the ALMA observations (PA$=157.3\pm0.3^{\circ}$). The surface brightness of our model measured at 5\arcsec separation along the PA (i.e., at the location of the arcs) is $\sim3~\mu$Jy arcsec$^{-2}$, in agreement with the contrast per resolution element reported in Figure 10 of \citet{schneider2014}, for which a resolution element was defined as a 0.1\arcsec wide photometric aperture. Most notably, our best fit model has an inclination of $i=88.3^{\circ}$ from face-on, which disagrees substantially from the ALMA observations ($i=56.2\pm0.4^{\circ}$)\citep{macgregor2022}. The scale height of our best fit model is $h=0.35$, which is extremely large and incompatible with the ALMA constraint ($h=0.04\pm0.02$) \citep{macgregor2022}. Figure \ref{fig:posteriors} shows the posterior distributions of our model fits. 

The middle panel of Figure \ref{fig:bestfitmodel} shows the residuals of our data set after subtracting the best fit model. The right panel shows those same residuals with our best fit model, rotated by $\Delta$PA$=90^{\circ}$, injected \emph{prior} to our KLIP data reduction. This panel clearly shows our disk model to the NE and SW, where we would expect to see the minor axis of the disk. We can therefore conclude that the KLIP reduction method is not over-subtracting the disk along the minor axis and cannot explain the observed absence of a forward scattering side of the disk.

Given the very large discrepancy in inclination with the ALMA data, we find our best fit model shown in Figure \ref{fig:bestfitmodel} to be unreliable. Qualitatively, our fitting routine is attempting to reproduce the seemingly ``detached" arcs of material to the NW and SE via an edge-on disk orientation. This circumvents the issue regarding an absence of forward scattering along the minor axis of a semi-inclined disk. The azimuthal extent of the arcs causes the fitting routine to adopt a large scale height. This behavior of our fitting routine again points to a lack of forward-scattering material along the minor axis of the ALMA-ring. 

In an attempt to produce a more physically plausible model, we restricted our disk model to the same orientation as measured by the ALMA observations. Specifically, we set the PA$=157.3^{\circ}$ and inclination $i=56.2^{\circ}$, and let all other parameters vary freely according to Table \ref{table:mcmc_parameters}. We adopted 14 walkers, 14k MCMC steps, and 7k burn-in steps for our MCMC analysis.

\begin{figure}[h]
	\centering	
	\includegraphics[width=7in]{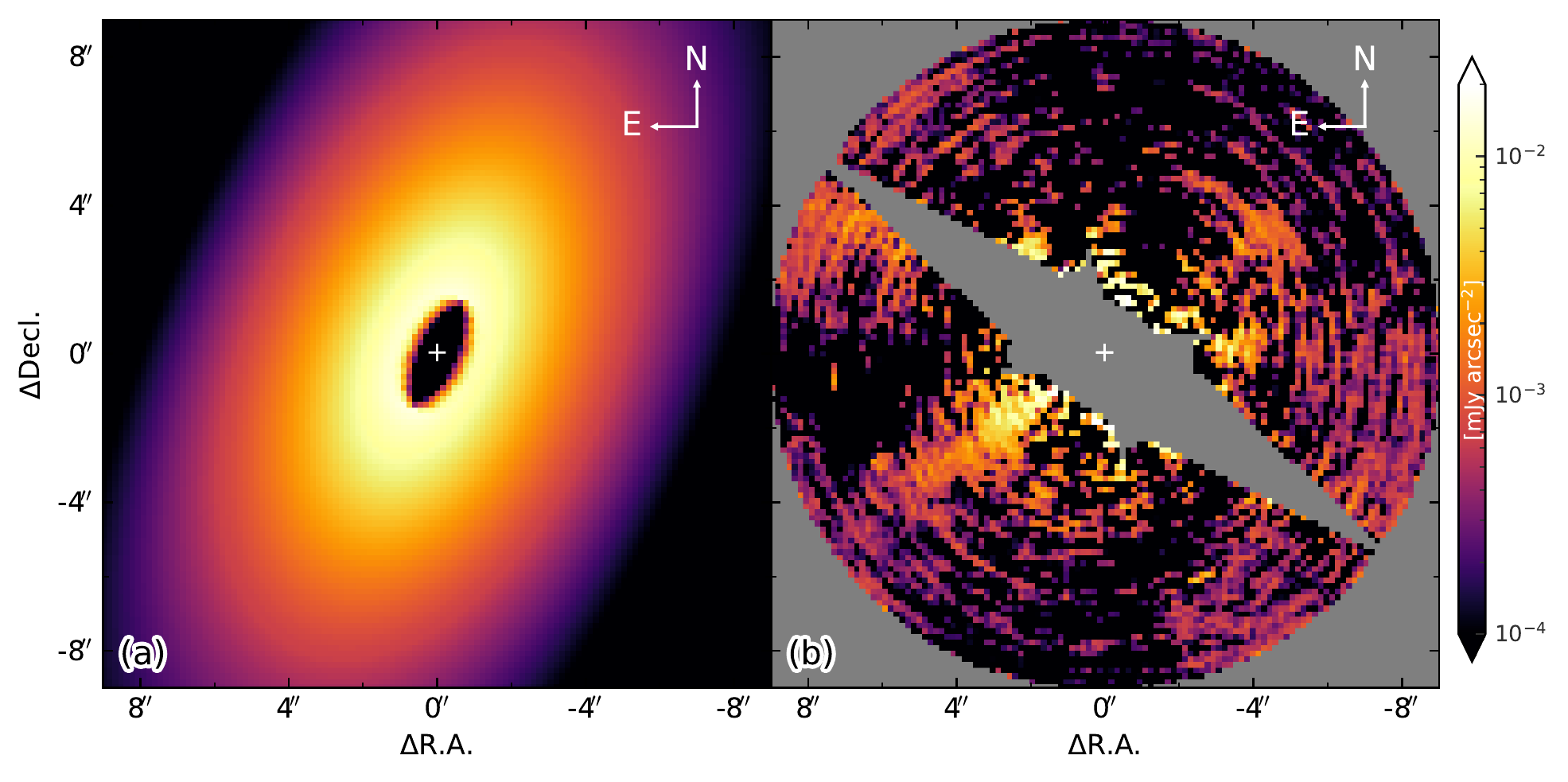}
	\caption{\emph{Left:} Best fit model to the observations while restricting the PA and inclination of the disk to the ALMA-measured values. \emph{Right:} KLIP-reduced data set with the best fit model subtracted showing significant correlated residuals. \label{fig:bestfitalmamodel}}
\end{figure}

\begin{figure}[h]
	\centering	
	\includegraphics[width=7in]{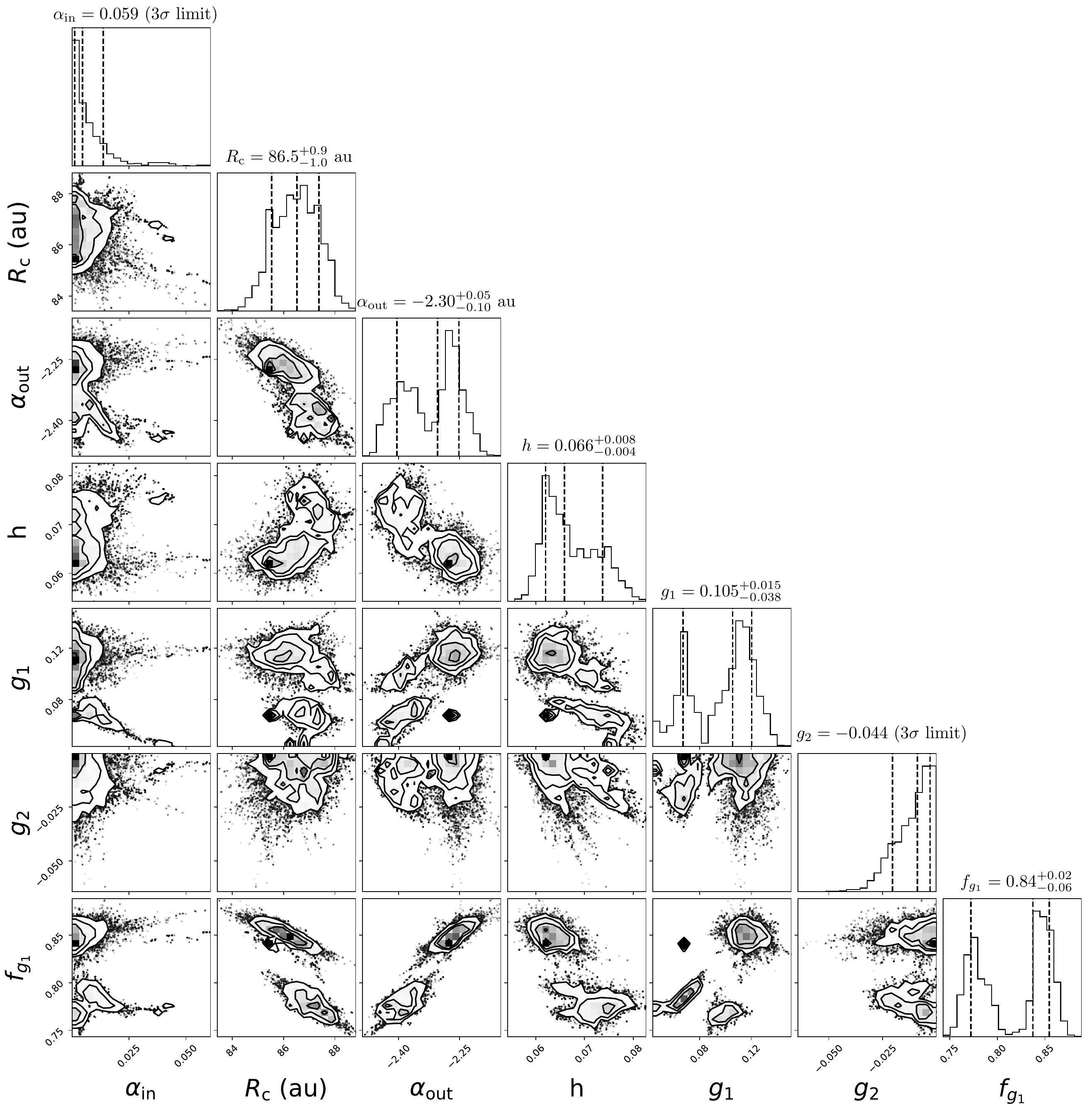}
	\caption{Posterior distribution for our disk model constrained to the orientation measured by ALMA.\label{fig:almaposteriors}}
\end{figure}

The left panel of Figure \ref{fig:bestfitalmamodel} shows the best fit disk model. The right panel shows the residuals after disk subtraction, revealing correlated residuals in excess of those of our best fit model shown in Figure \ref{fig:bestfitmodel}. Figure \ref{fig:almaposteriors} shows the posterior distribution for our best fit parameters when constrained to the ALMA-measured orientation. The best fit HG scattering asymmetry parameters are $g_1 \sim 0.1$ and $g_2 \sim 0.0$, indicating relatively isotropically scattering dust grains. The scale height $h=0.07^{+0.01}_{-0.00}$ and radius of the peak density $R_C=86.5^{+0.9}_{-1.0}$ AU of our model are both in near agreement with the symmetric model fit by \citet{macgregor2022} to ALMA data. The inner power law $\alpha_{\rm in}=0.00^{+0.01}_{-0.00}$ suggests material extends interior to the outer arcs, similar to the excess flux imaged by \citet{schneider2014}. 

We note that for these fits, we ignored eccentricity and assumed the disk was centered on the star. We attempted to include eccentricity, to first order, by simply offsetting the model from the star. All MCMC-retrieved centers were shifted to unrealistic values inconsistent with ALMA observations and we consider these models unreliable.

\section{Discussion} \label{sec:discussion}

\subsection{The star}

\subsubsection{PSF mis-match \& variability\label{psfvariability}}
In spite of a smaller absolute $\Delta(B-V)$ than previous observations, our reference star, HD 58895, provided a poor match to the HD 53143 PSF. Upon closer inspection, the PSF mis-match was determined to be worst during the last two epochs during which we observed HD 53143. We observed no variability in the reference star PSF, indicating that the variability in the PSF fitting was likely due to HD 53143.

\begin{figure}[h]
	\centering	
	\includegraphics[width=4in]{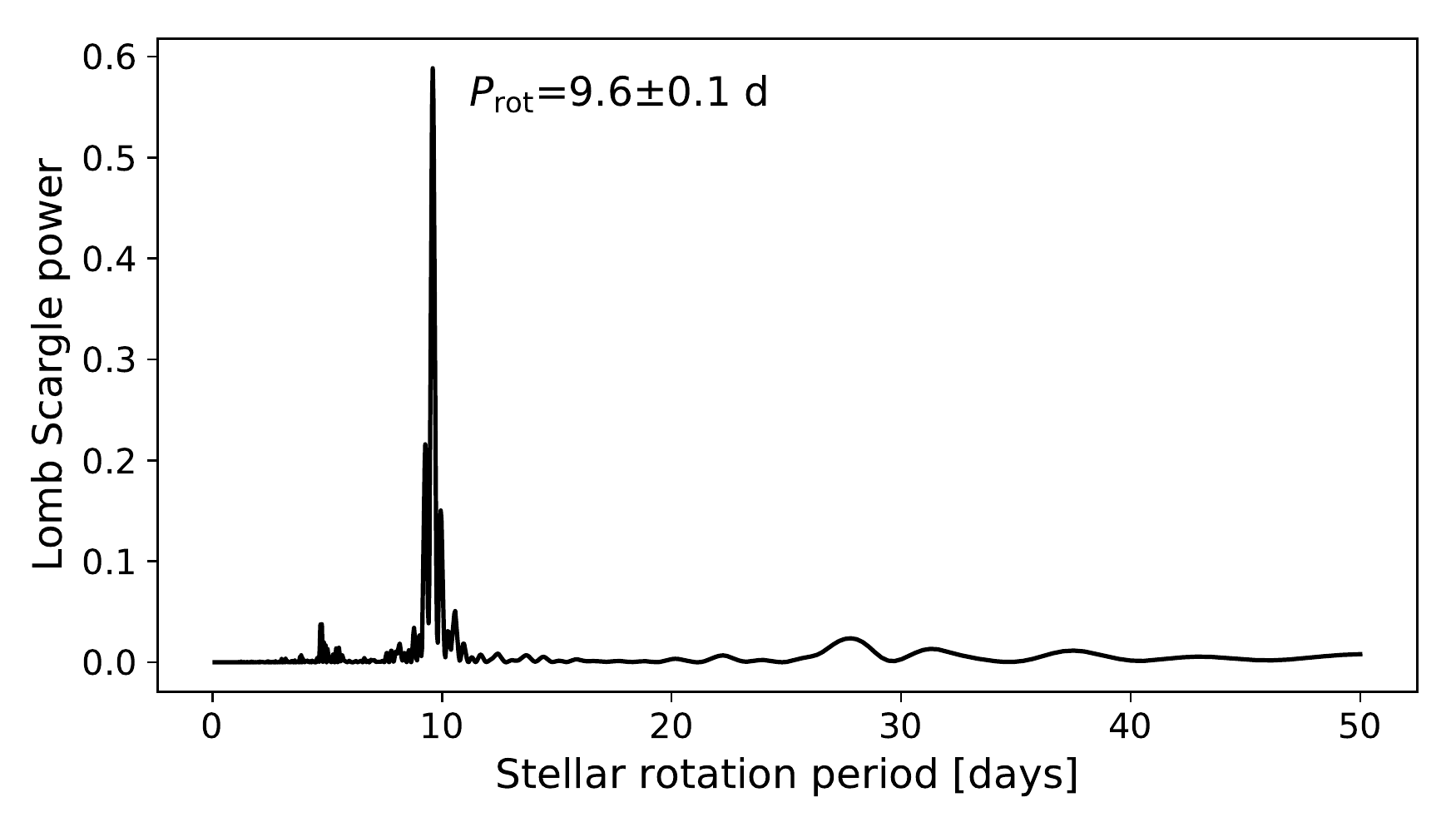}
	\caption{Lomb Scargle periodogram of the Cycle 3 TESS data showing the 9.6$\pm$0.1 d rotation period of HD 53143. This period is typical of K0 dwarfs of ~1 Gyr in age.\label{fig:periodogram}}
\end{figure}

In an effort to better understand the nature of the HD 53143 star, we analyzed its TESS lightcuves. HD 53143 is located near the TESS continuous viewing zone, providing an extended temporal baseline. We downloaded calibrated, two minute cadence TESS light curves of HD 53143, spanning 24 sectors between Sector 1 and Sector 39, from the Mikulski Archive for Space Telescopes (MAST)\footnote{https://mast.stsci.edu. The specific observations analyzed can be accessed via \dataset[DOI: 10.17909/t9-nmc8-f686]{https://doi.org/10.17909/t9-nmc8-f686}}. We chose to download Simple Aperture Photometry (SAP) light curves rather than Pre-search Data Conditioning (PDC) light curves to avoid removing the rotational variability. The light curves showed signs of significant spot evolution, with changes to the amplitude and phase of rotational variability occurring between Cycle 1 and Cycle 3. The final four months of TESS observations were obtained from March to June of 2021, prior to our GO 16202 STIS observations in October of 2021. These observations exhibit a sinusoidal rotation signal, indicative of the presence of a stable spot group. We measured the variability period during Cycle 3 to be 9.6$\pm$0.1 d using a Lomb Scargle periodogram as shown in Figure \ref{fig:periodogram}. 

\begin{figure}[h]
	\centering	
	\includegraphics[width=4in]{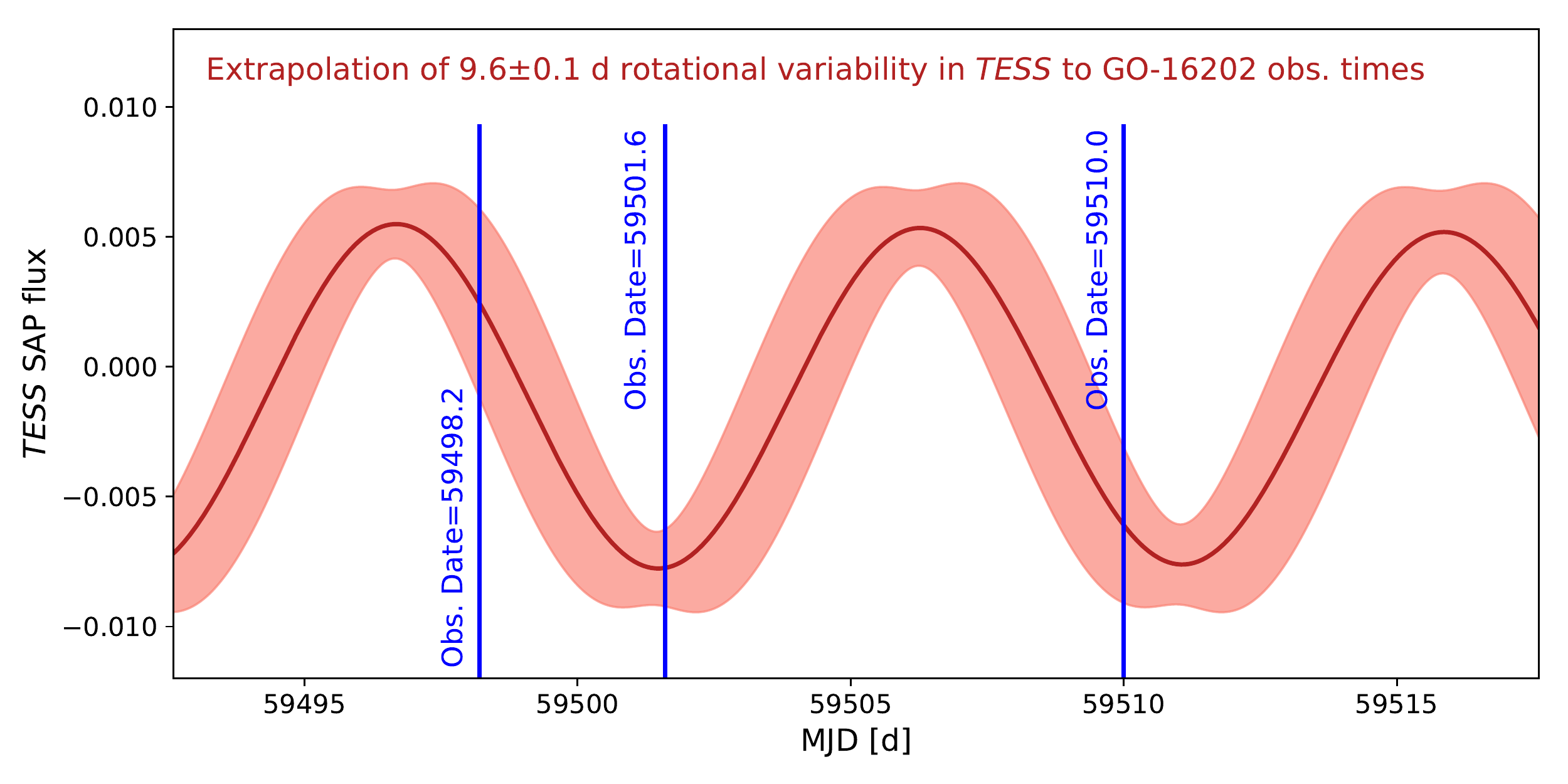}
	\caption{Predicted rotational variability during the Oct 2021 HST STIS observations. The HST observations taken near peak stellar brightness and minimum brightness correspond with over and under-subtractions in the HST data. The rotational variability is shown in red and is extrapolated from the rotation period and TESS light curve using bootstrap fits to the stable sinusoidal modulation in the light curve from Mar-Jun 2021.\label{fig:tessextrap}}
\end{figure}

We used a bootstrap approach to extrapolate the sinusoidal variability of the TESS data from Mar-Jun 2021 to our STIS observation window in Oct 2021. We assumed the dominant spot persisted over the four months after the end of the TESS window, which is supported by the presence of consistent patterns of rotational variability that last similar amounts of time in the rest of the multi-year TESS light curve. For each of 1000 trials, we randomly dropped 10\% of the light curve during the Mar-Jun 2021 observations and fit a sine with a period sampled from the 1-$\sigma$ range of a normal distribution centered at 9.6 days and a width of 0.1 days. We computed the average and standard deviation of the predicted flux values across all trials at each time during the GO 16202 window. We estimate the flux of HD 53143 to be significantly lower during the last two epochs of GO 16202 STIS observations, as shown in Figure \ref{fig:tessextrap}. This qualitatively agrees with the behavior of our STIS PSF reference fits---our reference star HD 58895, which is bluer than HD 53143, should provide a worse fit when HD 53143 appears redder due to enhanced star spots, which corresponds to the minima of the lightcurve shown in Figure \ref{fig:tessextrap}. We conclude that HD 53143 may exhibit star spots at a level that significantly impact STIS coronagraphic PSF subtraction.

This behavior was not noticed during the two epochs of STIS observations obtained during 2011 as part of the GO 12228 program \citep{schneider2014}. While spot evolution timescales prohibit us from extrapolating the TESS data obtained seven years after the GO 12228 observations (Giles et al. 2017), the precision of the TESS rotation period is sufficient to constrain the \emph{relative} phase during the 2011 observations. We performed 100,000 Monte Carlo trials in which the stellar rotation period is drawn from the same period distribution as in our previous model. For each period, we folded the two GO 12228 times in phase and compare the magnitude of the phase difference. We found a mean and standard deviation in phase of 0.25$\pm$0.1, where phase is normalized from zero to one. We may therefore say that the GO 12228 observations were not taken at opposite rotational phases and likely occurred at similarly high or low flux values, consistent with the lack of any observed changes to the quality of the reference PSF during the GO 12228 observations.

\subsubsection{Age}

Given the significant observed stellar variability and massive debris disk of this reportedly Sun-like star, we revisited the age estimates for HD 53143. We can attempt to estimate the age of the star from membership in a  young association, gyrochronology, and chromospheric emission. Although membership in IC2391 has been claimed, Banyan $\Sigma$ \citep{gagne2018} finds a 0\% probability of membership with IC2391 or any other young stellar association using Gaia DR2 parallaxes and proper motions.

The publicly available HARPS radial velocity database HARPS-RVBANK\footnote{https://www2.mpia-hd.mpg.de/homes/trifonov/HARPS\_RVBank.html. Note HD 53143 star is referred to as GJ260.} \citep{trifonov2020} has 26 measurements of HD 53143, which show clear variability with an amplitude of $\sim$37~m~s$^{-1}$. The periodogram calculated on that site produces the strongest period at 9.6 d.  This agrees with the stellar rotation period measured from TESS data, indicating that the RV signal is due to spots. We apply gyrochronology to estimate the age of the star using the calibration in \citet{mamajek2008} and the $B-V$ color of the star.  The star is spectral type G9-K0 with a $B-V$=0.77 - 0.81 depending on the source (e.g., Tycho-2, \cite{hog2000} and \cite{koen2010}), which yields an age of 540 Myr (460-600 Myr). We also applied the stardate code that combines isochrone and gyrochronology \citep{angus2019} using the stellar parameters from Gaia DR3 (Teff=5332, logg=4.45) and got an age of $1.29^{+1.01}_{-0.68}$ Gyr.

Another constraint on the stellar age comes from stellar activity.  The predicted Ca chromospheric activity index R'(HK) at an age of 540 Myr in \citet{mamajek2008} is about -4.4.  The measured R'(HK) for this star is -4.52 $\pm$ 0.05\citep{herrero2012,borosaikia2018}, which would make it a bit older.  \citet{stanfordmoore2020} use the star's B-V color and R'(HK) index to find an age between 245 Myr and 9.24 Gyr, with a median posterior value of 1.31 Gyr.

Altogether, the best constraint we can place is that the HD 53143 star is between 500 Myr and 2.4 Gyr in age.

\subsubsection{Inclination}

The derived stellar inclination is sensitive to the assumed stellar radius and to the assumed uncertainties. Taking stellar radius $R_{\star}=0.90\pm 0.02$ $R_{\sun}$ \citep{gaiamission,gaiadr3}, the observed $9.6 \pm 0.1$ day rotational period, and vsini$=4.0 \pm 0.7$ \citep{saar1997,nordstrom2004} we find a stellar inclination of $54 \pm 8^{\circ}$. This agrees well with the inclination of the disk seen by ALMA. Using somewhat different values, \citet{hurt2022} found a stellar inclination of $67\pm18^{\circ}$, which is still consistent with the ALMA disk given the larger uncertainties.

\subsubsection{Binarity}

We find no evidence for a stellar companion that would affect the dynamics of the observed disk around HD 53143. A query for common proper motion companions reveals a low-mass $\sim0.1$ $M_{\sun}$ star with similar proper motion at 28 kAU \citep{kervella2019}. The \emph{Gaia} EDR3 shows no evidence of a proper motion anomaly with residuals $<4$ m/s \citep{gaiadr3}, and no additional radial velocity signals were noticed in the HARPS dataset.

\subsection{Outer disk}

The multi-wavelength data available for the HD 53143 disk seems to suggest two possible scenarios for the outer disk at $\sim$90 AU. The first is that the small dust grains observed with HST STIS are smoothly distributed over a disk with the same orientation as their larger counterparts observed with ALMA. This azimuthally symmetric disk would require the small dust grains to be roughly isotropically scattering over the range of observable scattering angles, unlike most other observed debris disks. This may suggest that we are observing grains much smaller than the wavelength. Our fits indicate that in this scenario, dust extends into the inner regions of the system.

Truly isotropically scattering dust is not the only explanation. The scattering phase function must only \emph{appear isotropic over the range of observable scattering angles}. This could actually be explained by larger dust that is extremely forward scattering, but with a phase function that is relatively flat over the observed angles. Such a phase function would likely be even more forward scattering than that observed by \citet{hedman2015} and would imply a minimum grain size significantly larger than the wavelength.

An alternative explanation for the apparent absence of forward scattering could simply be that there is less dust along the minor axes, or equivalently, density enhancements at the ansae, which we do not explicitly model here. Density enhancements seen at visible wavelengths could result from an unseen companion trapping inward-migrating small dust grains into exterior mean motion resonances. We consider this unlikely, as the drag time is much longer than the collision time for micron sized grains at these circumstellar distances \citep{stark2009}. However, density enhancements could also result from increased \emph{production} of small grains at the ansae. Increased production of small grains could be due to planetesimals trapped in the exterior mean motion resonances of a planet \citep{wyatt2006}. Alternatively, given that ALMA observations constrain the apastron of the HD 53143 outer debris ring to be near the NW ansa \citep{macgregor2022}, increased production of small grains could be due to increased collision rates of the eccentric parent bodies at periastron and apastron. In both of these scenarios we would expect density enhancements in the ALMA data set. There is some evidence for a density enhancement along the NW ansa in the ALMA observations, which has been attributed to an ``apocenter glow" effect expected for eccentric planetesimal belts \citep{pan2016}. 

The simulations of \citet{lee2016} suggest that small dust grains originating from an eccentric belt should create an asymmetric halo beyond the belt, with an extension in the direction of apastron. The degree of this asymmetry depends on where dust grains are launched from in the belt; preferential production of dust grains at the parent bodies' periastron produces a larger halo asymmetry, while preferential production at apastron mutes the asymmetry. We examined radial cuts through our STIS observations along the disk's PA. Our analysis showed no significant flux asymmetry between the NW and SE sides of the disk beyond 5\arcsec, potentially favoring the apastron production scenario. 

\subsection{Inner disks}

\citet{chen2014} modeled the HD 53143 IR SED using \emph{WISE} and \emph{Spitzer} MIPS photometry, as well as IRS spectra, and reported flux in excess of the stellar photosphere corresponding to dust in the 0.6 -- 16 AU region. \cite{schneider2014} directly imaged excess flux extending further out, from $\sim5.5$--$55$ AU. The imaged flux was azimuthally symmetric in nature, suggesting a possible disk with a face-on orientation. \cite{macgregor2022} confirmed the presence of warm dust with the detection of a significant excess in ALMA Band 6 and estimated the circumstellar distance to be $\sim$25 AU with a width of $\sim5$ AU. However, the ALMA data has a bi-lobed appearance, which is more consistent with an edge-on inner disk.

Unlike other observed debris rings that feature sharp inner edges \citep[e.g.,][]{kalas2005,schneider2014,perrin2015,millarblanchaer2016}, our ALMA-constrained model fit to the outer disk has a very shallow inner radial power law of $\alpha_{\rm in}=0.0$. Notably, this is even \emph{less} steep than the \emph{outer} halo component ($\alpha_{\rm out}=-2.3$). This suggests that dust from the outer disk may be migrating in to the inner regions of the system, possibly contributing to the observed warm dust. Assuming that the optical depth of the HD 53143 disk can be approximated as $\tau\sim L_{\rm IR}/L_{\star}$, this seems unlikely for a disk with optical depth $\tau\sim10^{-4}$ based on the simulations of \cite{stark2009}. Alternatively, our shallow inner radial power law may also be due to an extended halo from a dynamically separate inner disk.
	
\section{Conclusions} \label{sec:conclusions}

The HD 53143 debris disk has a peculiar appearance that may suggest a dynamically active system. This disk lacks any clear sign of a forward scattering side, as expected near the minor axis of the disk, connecting two ``arcs" of cold material at $\sim$90 AU. Symmetric disk model fits to our HST STIS data set avoid placing disk flux in the region where the minor axis is expected to be, resulting in unphysical edge-on models. Models with constrained geometries provide poorer fits and require either near-isotropically scattering small dust, or very forward scattering (i.e., large) dust grains whose phase function appears relative flat over all observable scattering angles. Alternatively, the apparent absence of forward scattering could be explained by a lack of dust along the minor axes, with the ansae of the HD 53143 disk being regions of enhanced grain production due to the influence of an unseen planet. Observations suggest the presence of multiple inner disks, the geometry of which are not well understood. The apparent dichotomy between the dynamical distributions of small and large grains in this system suggests that the HD 53143 disk may provide insights into the dynamics of dust in the outer regions of planetary systems.

\acknowledgments

The authors acknowledge Pierre Kervella for his assistance assessing the binarity of HD 53143. This research is based on observations made with the NASA/ESA Hubble Space Telescope obtained from the Space Telescope Science Institute, which is operated by the Association of Universities for Research in Astronomy, Inc., under NASA contract NAS 5–26555. These observations are associated with program 16202. Part of the computations presented here were conducted in the Resnick High Performance Computing Center, a facility supported by Resnick Sustainability Institute at the California Institute of Technology. This work has made use of data from the European Space Agency (ESA) mission {\it Gaia} (\url{https://www.cosmos.esa.int/gaia}), processed by the {\it Gaia} Data Processing and Analysis Consortium (DPAC, \url{https://www.cosmos.esa.int/web/gaia/dpac/consortium}). Funding for the DPAC has been provided by national institutions, in particular the institutions participating in the {\it Gaia} Multilateral Agreement. E.C. acknowledges funding from the European Research Council (ERC) under the European Union's Horizon Europe research and innovation programme (ESCAPE, grant agreement No 101044152).

 \bibliography{ms_v3.bbl}
\bibliographystyle{aasjournal}

\end{document}